\documentclass[prb,twocolumn,showpacs,superscriptaddress,floatfix, nofootinbib]{revtex4-2}
\usepackage{amsmath,amssymb,amsfonts,mathrsfs, amsbsy}
\usepackage{graphicx}
\usepackage[bookmarks=true, colorlinks=true, linkcolor=blue, urlcolor=blue, citecolor=blue, bookmarks=true, hyperindex=true]{hyperref}

\newcommand{\be}{\begin{equation}}
\newcommand{\ee}{\end{equation}}

\newcommand{\beq}{\begin{eqnarray}}
\newcommand{\eeq}{\end{eqnarray}}

\def\H1{\widehat{H}_1}

\newcommand{\ket}[1]{\left| #1 \right>}
\newcommand{\bra}[1]{\left< #1 \right|}
\newcommand{\bs}{\boldsymbol}

\begin{document}

\title{Adiabatic eigenstate deformations and weak integrability breaking of Heisenberg chain}

\author{Pavel Orlov}
\affiliation{Moscow Institute of Physics and Technology, Dolgoprudny, Moscow Region 141700, Russia}
\affiliation{Russian Quantum Center, Skolkovo, Moscow 143025, Russia}

\author{Anastasiia Tiutiakina}
\affiliation{Laboratoire de Physique Th\'{e}orique et Mod\'{e}lisation, CNRS UMR 8089,
	CY Cergy Paris Universit\'{e}, 95302 Cergy-Pontoise Cedex, France}
\affiliation{Russian Quantum Center, Skolkovo, Moscow 143025, Russia}

\author{Rustem Sharipov}
\affiliation{Physics Department, Faculty of Mathematics and Physics, University of Ljubljana,
Ljubljana, Slovenia}
\affiliation{Russian Quantum Center, Skolkovo, Moscow 143025, Russia}

\author{Elena Petrova}
\affiliation{Institue of Science and Technology Austria, Am Campus 1, 3400 Klosterneuburg, Austria}
\affiliation{Russian Quantum Center, Skolkovo, Moscow 143025, Russia}

\author{Vladimir Gritsev}
\affiliation{Institute for Theoretical Physics Amsterdam, University of Amsterdam, P.O. Box 94485, 1090 GL Amsterdam, The Netherlands}
\affiliation{Russian Quantum Center, Skolkovo, Moscow 143025, Russia}

\author{Denis\,V.~Kurlov}
\affiliation{Department of Physics, University of Basel, Klingelbergstrasse 81, CH-4056 Basel, Switzerland}
\affiliation{Russian Quantum Center, Skolkovo, Moscow 143025, Russia}
\affiliation{National University of Science and Technology ``MISIS'', Moscow 119049, Russia}

%\date{\today}

\begin{abstract}
We consider the spin-$\frac{1}{2}$ Heisenberg chain (XXX model) weakly perturbed away from integrability by an isotropic next-to-nearest neighbor exchange interaction. 
Recently, it was conjectured that this model possesses an infinite tower of {\it quasiconserved} integrals of motion (charges) [\href{https://doi.org/10.1103/PhysRevB.105.104302}{D. Kurlov {\it et al.}, Phys. Rev. B {\bf 105}, 104302 (2022)}].  
In this work we first test this conjecture by investigating how the norm of the adiabatic gauge potential (AGP) scales with the system size, which is known to be a remarkably accurate measure of chaos.  
We find that for the perturbed XXX chain the behavior of the AGP norm corresponds to neither an integrable nor a chaotic regime,
which supports the conjectured quasi-integrability of the model. 
We then prove the conjecture and explicitly construct the infinite set of quasiconserved charges. Our proof relies on the fact that the XXX chain perturbed by next-to-nearest exchange interaction can be viewed as a truncation of an integrable long-range deformation of the Heisenberg spin chain. 
\end{abstract}

\maketitle

\section{Introduction}

Quantum chaos has become a subject of intensive research over the last decades. 
Despite significant efforts and numerous milestones achieved, there still is a large number of open questions (for a review, see, e.g. Refs.~\cite{d2016quantum, haake1991quantum, stockmann_1999, Berry1989}).
On the contrary, in classical systems, chaos is a well understood phenomenon that relies on the exponential sensitivity of the phase space trajectories to initial conditions~\cite{vulpiani2009chaos}. 
This does not occur in integrable systems, because their trajectories are confined to certain subregions (tori) of the phase space, due to the presence of many conservation laws~\cite{Arnold1989}. Moreover, the renowned Kolmogorov-Arnold-Moser (KAM) theorem states that classical integrable systems under weak integrability-breaking perturbations remain stable for a sufficiently long time, because such perturbations do not destroy and only slightly deform most of the phase-space tori~\cite{kolmogorov1979preservation, kolmogorov1954conservation, moser1962invariant, arnol1963small, arnold1963proof}. 

Extending the physical picture of chaos from classical systems to the quantum ones is far from being straightforward, already due to the fact that the notion of phase space trajectories does not apply to quantum systems. Thus, in the quantum case one has to define chaos differently. A particularly successful and widely accepted approach to quantum chaos is based on the random matrix theory (RMT)~\cite{Brody1981, Guhr1998} and the celebrated eigenstate thermalization hypothesis (ETH)~\cite{Srednicki1994, DAlessio2016, Deutsch2018},
 which describes how isolated quantum systems relax locally to thermal equilibrium. In the context of RMT and ETH, chaotic quantum systems are most commonly characterized by their spectral properties, such as the level spacing statistics~\cite{Rabson2004, Rigol2010, Santos2010}, mean gap ratio~\cite{oganesyan2007localization, atas2013distribution}, and spectral form factor~\cite{brezin1997spectral, bertini2018exact}.
For instance, according to the Bohigas-Giannini-Schmit conjecture \cite{bohigas1984characterization} chaotic systems exhibit Wigner-Dyson level spacing statistics due to the repulsion between the energy levels. On the contrary, in integrable quantum systems the energy levels are uncorrelated, so that the corresponding level spacing statistics is Poissonian~\cite{Berry1977}. 
Intuitively this is can be understood from the fact that integrable systems possess a macroscopic number of local conserved quantities (charges) that commute with the Hamiltonian and one another, which is a widely accepted criterion for quantum integrability.
This is also the reason why quantum integrable systems do not follow the ETH. Instead, their thermalization is governed by the so-called generalized Gibbs ensemble (GGE) that takes into account the additional conservation laws~\cite{essler2016quench, Essler2022, Gopalakrishnan2023}.

Recently, an alternative approach to quantum chaos has been formulated, which utilizes concepts of quantum geometry~\cite{Kolodrubetz2013} and relies on the rate of deformations of eigenstates under infinitesimal perturbations.  
It turns out that the generator of these deformations, dubbed the adiabatic gauge potential (AGP), provides an exceptionally sensitive measure of chaos~\cite{pandey2020adiabatic, Sierant2019}.
Indeed, the Frobenius norm of the AGP is nothing other than the distance between the nearby eigenstates (the so-called Fubini-Studi metric)~\cite{Provost1980, Page1987, Kolodrubetz2017}. It can be easily shown that this norm scales with the system size in a drastically different manner for chaotic and integrable quantum systems. Namely, the AGP norm exhibits an exponential scaling for chaotic systems described by the ETH, whereas for integrable systems the scaling is polynomially bounded~\cite{pandey2020adiabatic}. Therefore, one can say that quantum chaos manifests itself in the exponential sensitivity of the eigenstates to the integrability-breaking perturbations, which provides a certain analogy with the classical chaos. 
The AGP norm is remarkably accurate in distinguishing the chaotic systems from the integrable ones, since it is sensitive to integrability-breaking perturbations that are exponentially small in the system size. This by far exceeds the sensitivity of standard probes of chaos, such as the spectral form factor or level statistics. 
Moreover, from the practical perspective, the difference between the polynomially bounded and exponential scaling of the AGP norm is easy to detect numerically, even for relatively small system sizes. 
These and other arguments demonstrate that the AGP norm is an extremely useful tool for detecting chaotic behavior in quantum many-body systems. Over the past few years the AGP based approach has lead to a number of important achievements and insights on quantum chaos~\cite{leblond2021universality} and quantum control~\cite{Sels2017, Hatomura2021, Hartmann2019}. It was also used in the context of the many-body localization~\cite{Suntajs2019, Panda2020, Luitz2020, Sierant2020, Abanin2021, sels2021dynamical, Morningstar2022, Sierant2022, Nandy2022}.

One of the most important questions in the field of quantum chaos is related to the fate of many-body systems under weak integrability-breaking perturbations.  Generalizing the KAM theorem to the quantum case is a long-standing problem. Despite recent findings demonstrating some progress in this direction~\cite{Brandino2015}, a complete understanding is missing. To some extent this can be explained by the fact that even the very definition of quantum integrability is subtle~\cite{Caux2011}. 
Extensive research shows that weakly-nonintegrable quantum systems do not thermalize for sufficiently long times $t_{\text{th}} \sim \lambda^{-2}$, where $\lambda \ll 1$ is the strength of the perturbations~\cite{stark2013kinetic, mallayya2019prethermalization}, as can be understood using the Pauli master equation and Fermi's golden rule-like arguments~\cite{mori2018thermalization}. At times smaller than~$t_{\text{th}}$, weakly-nonintegrable systems exhibit a different, the so-called prethermal behaviour at earlier times~\cite{Berges2004, Bertini2015, Langen2016, durnin2021nonequilibrium}. It is believed that the prethermal phase should be described by some effective GGE~\cite{fagotti2013reduced, pozsgay2013generalized}. 

Remarkably, in some cases the thermalization time turns out to be much larger than the naive~$t_{\text{th}} \sim \lambda^{-2}$ scaling~\cite{Abanin2017, mori2018thermalization}. For instance, the spin-$\frac{1}{2}$ isotropic Heisenberg chain (XXX model), weakly perturbed by an isotropic next-to-nearest neighbor exchange interaction was found to exhibit transport behavior consistent with the thermalization time $t_{\text{th}} \sim \lambda^{-4}$~\cite{jung2006transport}. 
This anomalously large thermalization time was attributed to the existence of an {\it approximate} integral of motion, conserved with the accuracy $O(\lambda^2)$.
% A natural question is then what are the charges that define this effective prethermal GGE. Since the effective description is only valid at times $t \lesssim \lambda^{-2}$, it is obvious that these charges can be only {\it approximately} conserved with the accuracy $O(\lambda^2)$. 
This question has been further addressed in Ref.~\cite{kurlov2021}, which explicitly constructed a few higher-order quasiconserved charges for the spin-$\frac{1}{2}$ XXX model, weakly perturbed by an isotropic next-to-nearest neighbor exchange interaction. In the same work it was conjectured that and one can construct as many quasiconserved charges for the perturbed XXX chain, as there are exactly conserved charges for the unperturbed integrable model (infinitely many in the thermodynamic limit). Similalry, a few first quasiconserved charges were constructed for an isotropic XY chain perturbed by a next-to-nearest neigbor XY-interaction~\cite{bahovadinov2022many}. 
There have been numerous studies of spectral properties in weakly-nonintegrable models, for instance showing  a crossover from Poissonian to Wigner-Dyson level statistics, see e.g. Refs.~\cite{Rabson2004, Scaramazza2016, szasz2021weak, McLoughlin2022, Bulchandani2022}.
However, since the AGP norm has proven to be much more sensitive and efficient in detecting quantum chaos, it is natural to ask whether it can provide further insight into the behavior of weakly-nonintegrable systems. 

The aim of this work is twofold. First, we investigate the perturbed XXX chain  using the AGP-based approach. 
Steps in this direction (albeit with a different logic -- see discussion in Sec.~\ref{sec_AGP_scaling}) were already performed in Ref.~\cite{pandey2020adiabatic}. There, it was observed that apart from the polynomially bounded and exponential scaling for integrable and chaotic systems, respectively, the AGP norm can exhibit yet another regime. Namely, for weakly-nonintegrable systems one has a sharp crossover between the regimes of a polynomially bounded and exponential scaling of the AGP norm.  The latter regime has been associated with the emergence of exponentially slow relaxation dynamics~\cite{pandey2020adiabatic}.
The crossover was found to occur at a critical perturbation strength that is exponentially small in the system size. This picture agrees with our findings presented in this paper. We find that the AGP norm for an integrable model (XXX chain) weakly perturbed by an integrability-breaking perturbation (isotropic next-to-nearest neighbor exchange interaction) exhibits the crossover between the polynomially bounded and exponential scaling at a critical perturbation strength, which is exponentially small in the length of the chain. The observed behavior of the AGP norm is distinct from both integrable and purely chaotic regimes. We argue that this strongly supports the conjecture on the quasi-integrability of the perturbed XXX model, put forward in Ref.~\cite{kurlov2021}.

In the second part of the paper we proceed with the analytic proof of this conjecture. 
We present an explicit construction of the infinite set of quasi-conserved charges for the isotropic Heisenberg Hamiltonian perturbed by a weak next-to-nearest neighbor exchange interaction. We employ the idea of integrability-preserving long-range deformations, introduced in Ref.~\cite{Bargheer2009}, and show its direct relation to the notion of AGP. Our proof of quasi-integrability relies on the fact that the perturbed XXX chain can be viewed as a truncation of an integrable spin chain with long-range interactions.

The rest of the paper is organized as follows.
In Sec.~\ref{sec_model} we introduce the perturbed XXX model and discuss the conjecture on its quasi-integrability~\cite{note_quasi_integrability}.
In Sec.~\ref{sec_AGP_scaling} we briefly review the notion of the AGP. We numerically investigate the scaling of the AGP norm for our model and demonstrate that the results are consistent with the conjectured quasi-integrability.
Then, in Sec.~\ref{sec_quasi-int} we proceed with the analytic proof of the conjecture and present an explicit construction of an infinite set of the quasiconserved charges for the perturbed XXX chain.
Finally, in Sec.~\ref{sec_conclusions} we discuss our results and conclude. For the sake of completeness the paper is supplemented with an appendix where we briefly discuss another weakly-nonintegrable model -- isotropic XY chain perturbed by the next-to-nearest neighbor XY interaction, studied in Ref.~\cite{bahovadinov2022many}. 

\section{Perturbed XXX model} \label{sec_model}

In this section, we introduce the model and briefly discuss the conjecture on its quasi-integrability, put forward in Ref.~\cite{kurlov2021}.
We consider the Hamiltonian 
\begin{equation} \label{H_tot}
    H(\lambda) = H_0 + V(\lambda), 
\end{equation}
where $H_0$ is the integrable part and $V(\lambda)$ is the perturbation. The real parameter $\lambda$ controls the perturbation strength and we assume~$\lambda \ll 1$, so that the perturbation is weak. The term~$H_0$ corresponds to the spin-$\frac{1}{2}$ isotropic Heisenberg chain and reads
\begin{equation} \label{H_XXX}
    H_0 = J \sum_{j} {\bs \sigma}_j \cdot {\bs \sigma}_{j+1},
\end{equation}
where ${\bs \sigma}_j = \{ \sigma_j^{x}, \sigma_j^{y}, \sigma_j^{z} \}$ is the vector of Pauli matrices, the dot denotes the scalar product, and $J$ is the exchange coupling constant. Hereinafter we work in units with $J=1$. The Hamiltonian~(\ref{H_XXX}) is integrable and can be solved exactly by the Bethe ansatz~\cite{Bethe1931, korepin1997quantum}. The model possesses an infinite number of conserved charges that commute with the Hamiltonian and one another
\begin{equation}
    [H_0, {\cal Q}_n] = [{\cal Q}_m, {\cal Q}_n] = 0.
\end{equation}
We stress that the conserved charges~${\cal Q}_n$ are required to be local in the sense that they are given by a sum of operators with finite support. 
By convention, ${\cal Q}_1$ is the total magnetization, which is clearly conserved by $H_0$, and the second conserved charge coincides with the Hamiltonian itself, ${\cal Q}_2 = H_0$. The higher charges can be iteratively generated from ${\cal Q}_2$ as~\cite{Grabowski1994, grabowski1995structure, Grabowski1996}
\begin{equation} \label{XXX_charges_boost}
    {\cal Q}_{n+1} = \bigl[ {\cal B}[{\cal Q}_2], {\cal Q}_n \bigr],
\end{equation}
where ${\cal B}[{\cal Q}_2]$ is the so-called boost operator. Explicitly it is given by
\begin{equation}
    {\cal B}[{\cal Q}_2] =  \frac{1}{2i} \sum_{j} j {\bs \sigma}_j \cdot {\bs \sigma}_{j+1},
\end{equation}
 and we see that~${\cal B}[{\cal Q}_2]$ is constructed out of the second charge (Hamiltonian), cf. Eq.~(\ref{H_XXX}). Note that every next charge has a larger support as compared to the previous one. For the Heisenberg model the $n$th charge ${\cal Q}_n$ generated by Eq.~(\ref{XXX_charges_boost}) is a sum of term with the support on up to~$n$ lattice sites. In addition, ${\cal Q}_n$ usually contains terms that are also present in the previous charges~\cite{Grabowski1994}. It is convenient to work with a different basis~$\{ Q_m^{(0)} \}$ in which every next charge does not contain the terms from the previous charges. The first charges are the same in both bases (${\cal Q}_m = Q_m^{(0)}$ for $m=1,2$), whereas the next two conserved charges in this basis read
 \begin{align} 
     Q_3^{(0)} &= \sum_{j}\left( {\bs \sigma}_j \times {\bs \sigma}_{j+1} \right) \cdot {\bs \sigma}_{j+2},  \label{Q_3_XXX}\\
     Q_4^{(0)} &= \sum_{j} \bigl( \left( {\bs \sigma}_j \times {\bs \sigma}_{j+1} \right) \times {\bs \sigma}_{j+2} \bigr) \cdot {\bs \sigma}_{j+3} + \sum_{j} {\bs \sigma}_{j} \cdot {\bs \sigma}_{j+2},  \label{Q_4_XXX}
 \end{align}
where the cross denotes the vector product and the general form of $Q_n^{(0)}$ for the XXX model can be found in~Ref.~\cite{Grabowski1994}.

Let us now turn to the second term in Eq.~(\ref{H_tot}).
Following Ref.~\cite{kurlov2021}, for the perturbation~$V(\lambda)$ we take the isotropic next-to-nearest neighbor exchange interaction:
\begin{equation} \label{NNN_exchange}
    V(\lambda) = \lambda \sum_{j} {\bs \sigma}_j \cdot {\bs \sigma}_{j+2}.
\end{equation}
Note that both $H_0$ and $V(\lambda)$ are translation and $SU(2)$ invariant and we assume that the system is in the thermodynamic limit. 
The perturbation~(\ref{NNN_exchange}) breaks the integrability and the quantities~$Q_n^{(0)}$ are no longer conserved since they do not commute with the total Hamiltonian~$H(\lambda)$. 
Moreover, one clearly has $\| [H(\lambda), Q_n^{(0)}] \| \propto \lambda$, so that the quantities~$Q_n^{(0)}$ change significantly over times much shorter than $t_{\text{th}}$. Thus, they can not be responsible for the existence of the prethermal phase. One can try to deform~$Q_n^{(0)}$ into 
\begin{equation} \label{quasi_charge_gen}
    \tilde Q_n(\lambda) = Q_n^{(0)} + \lambda \, Q^{(1)}_n,
\end{equation}
where the correction~$Q_n^{(1)}$ is chosen such that the deformed charges $\tilde Q_n(\lambda)$ satisfy
\begin{equation}
    \| [ H(\lambda), \tilde Q_n (\lambda) ] \| \propto \lambda^2,
\end{equation}
and also commute with each other with the accuracy $O(\lambda^2)$. If this is possible, then the {\it quasi-conserved} charges~$\tilde Q_n(\lambda)$ constraint the dynamics and prevent the system from being truly chaotic during the prethemral phase.
With a tedious but straightforward brute force approach the authors of Ref.~\cite{kurlov2021} have constructed the first four nontrivial quasi-conserved charges, $\tilde Q_n(\lambda)$ with $3 \leq n \leq 6$, for the perturbed Heisenberg model. For instance, the correction to $Q_3^{(0)}$ was found to be
\begin{equation}
    Q_3^{(1)} = \sum_{j}\left( {\bs \sigma}_j \times {\bs \sigma}_{j+1} \right) \cdot {\bs \sigma}_{j+3} + \sum_{j}\left( {\bs \sigma}_j \times {\bs \sigma}_{j+2} \right) \cdot {\bs \sigma}_{j+3}.  
\end{equation}
It was then conjectured that one has as many quasi-conserved charges for the perturbed model as there are exact conservation laws for the unperturbed one.
We prove this conjecture in Sec.~\ref{sec_quasi-int}, where we explicitly construct an infinite tower on quasi-conserved charges. However, before doing so, let us first test the conjecture of Ref.~\cite{kurlov2021} using an extremely sensitive probe of chaos -- the AGP norm.

\section{Adiabatic gauge potential and integrability breaking} \label{sec_AGP_scaling}

In this section we briefly review the notion of adiabatic gauge potential and its power in detecting chaos. For an in-depth discussion see Refs.~\cite{pandey2020adiabatic} and~\cite{Kolodrubetz2017}. We then numerically investigate the scaling of AGP norm with the system size for the perturbed Heisenberg model. We show that the results are consistent with the conjectured quasi-integrability of the model.

\subsection{Adiabatic gauge potential}

Consider a Hamiltonian~$H(\lambda)$ depending on a parameter~$\lambda$. Let $\{ \ket{n(\lambda)} \}$ be its orthonormal eigenbasis, so that one has
\begin{equation} \label{Schroedinger_eq}
    H(\lambda) \ket{n(\lambda)} = E_n(\lambda) \ket{n(\lambda)}.
\end{equation}
Then, there exists a unitary transformation that adiabatically rotates the eigenstates as 
\begin{equation} \label{basis_rotation_U}
    \ket{n(\lambda)} = U(\lambda) \ket{n_{0}},
\end{equation}
where $\ket{n_{0}} = \ket{n(0)}$. The generator of this transformation is the so-called adiabatic gauge potential defined as
\begin{equation} \label{AGP_unitary}
    {\cal A}_{\lambda} = i \left[ \partial_{\lambda} U(\lambda) \right] U^{\dag}(\lambda),
\end{equation}
so that its action on the eigenstates is ${\cal A}_{\lambda} \ket{n(\lambda)} = i \partial_{\lambda} \ket{n(\lambda)}$.
It can be easily shown that the AGP satisfies the following operator equation~\cite{Jarzynski2013, Kolodrubetz2017}
\begin{equation} \label{AGP_op_rel}
    i \partial_{\lambda} H(\lambda) = \bigl[ {\cal A}_{\lambda}, H(\lambda) \bigr] - i {\cal F}(\lambda),
\end{equation}
where the operator ${\cal F}(\lambda)$ is diagonal in the eigenbasis of $H(\lambda)$. Explicitly, it is given by
\begin{equation} \label{F_op}
    {\cal F}(\lambda) = - \sum_{n} \frac{\partial E_n(\lambda)}{\partial \lambda} \ket{n(\lambda)} \! \bra{n(\lambda)}.
\end{equation}
The relation~(\ref{AGP_op_rel}) can be easily derived from the fact that the rotated Hamiltonian~$\tilde H(\lambda) = U^{\dag}(\lambda) H(\lambda) U(\lambda)$ commutes with its derivative $\partial_{\lambda} \tilde H(\lambda)$.
Similarly, Eq.~(\ref{F_op}) follows immediately from the Schr\"odinger equation~(\ref{Schroedinger_eq}). Differentiating both sides of Eq.~(\ref{Schroedinger_eq}) with respect to $\lambda$ and substituting $\partial_{\lambda}H(\lambda)$ from Eq.~(\ref{AGP_op_rel}), one arrives at~Eq.~(\ref{F_op}).

With the help of the Hellmann-Feynman theorem one can easily see that the matrix elements of the AGP calculated between the eigenstates of~$H(\lambda)$ read 
\begin{equation}
    \bra{m(\lambda)} {\cal A}_{\lambda} \ket{n(\lambda)} = -\frac{i}{\omega_{m n}} \bra{m(\lambda)} \partial_{\lambda} H(\lambda) \ket{n(\lambda)},
\end{equation}
where $\omega_{mn} = E_m(\lambda) - E_n(\lambda)$. In order to allow for degeneracies (accidental or not), one has to regularize the AGP as
\begin{equation} \label{AGP_reg}
    \bra{m} {\cal A}_{\lambda}(\mu) \ket{n} = -\frac{i \, \omega_{m n} }{\omega_{m n}^2 + \mu^2} \bra{m} \partial_{\lambda} H(\lambda) \ket{n},
\end{equation}
where $\mu$ is a small energy cutoff and we suppressed the dependence on $\lambda$ for brevity.
The results of Ref.~\cite{pandey2020adiabatic} show that the optimal cutoff choice is $\mu \sim L {\cal D}^{-1}$.
In order to lighten the notations, from now on we drop the argument~$\mu$, so that ${\cal A}_{\lambda}$ refers to the regularized AGP.

The Frobenius norm of the regularized AGP reads
\begin{equation} \label{AGP_norm}
    \| {\cal A}_{\lambda} \|^2 = \frac{1}{{\cal D}} \sum_{n}\sum_{m \neq n} | \bra{m} {\cal A}_{\lambda} \ket{n} |^2,
\end{equation}
where ${\cal D}$ is the dimension of the Hilbert space.
It is then straightforward to see that for chaotic systems, described by the ETH, the AGP norm scales exponentially with the system size.
Indeed, according to the ETH, the off-diagonal matrix elements of any local operator scale as $e^{-S/2}$, where $S$ is the entropy of the system~\cite{Srednicki1994}. Similarly, for the level spacings one has $\omega_{mn} \sim e^{-S}$. 
Thus, provided that the cutoff is chosen as $\mu \sim e^{-S}$, we immediately see that for chaotic systems $\| {\cal A}_{\lambda} \|^2 \sim e^{\kappa L}$, with some $\kappa >0$. 
On the contrary, for integrable models the AGP norm behaves differently and its scaling of the system size is bounded polynomially, as was demonstrated in Ref.~\cite{pandey2020adiabatic}. 
It turns out that for {\it weakly} nonintegrable systems the AGP norm exhibits yet another behavior, as we demonstrate below.

\subsection{Scaling of the AGP norm for the perturbed XXX model}

\begin{figure}[t!]
	\includegraphics[width=\columnwidth]{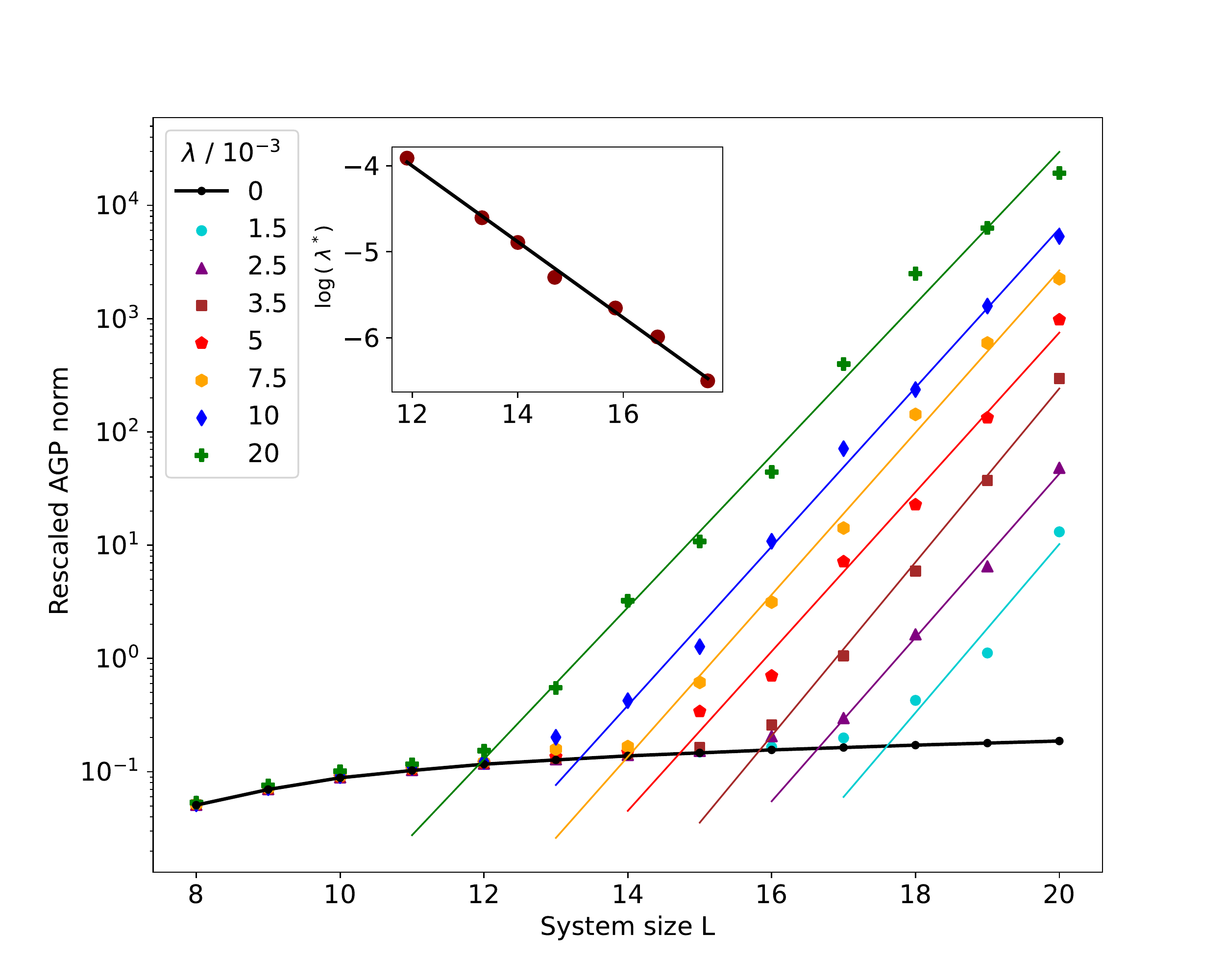}
    \caption{Main panel: The rescaled AGP norm $\|{\cal A}_{\lambda}\|^2/L$ as a function of the system size $L$  for the XXX Hamiltonian~$H_0$ in Eq.~(\ref{H_XXX}), weakly perturbed by the isotropic next-to-nearest exchange interaction~$V(\lambda)$, given by Eq.~(\ref{NNN_exchange}).
    The AGP is calculated using Eq.~(\ref{AGP_norm}). One can clearly see a crossover from the polynomially bounded to the exponential scaling, which occurs at a system-size dependent critical perturbation strength $\lambda^*(L)$. The solid lines are the exponential fits $\| {\cal A}_{\lambda} \|^2/L \propto e^{\kappa L}$, with $\kappa=1.62 \pm 0.05$. Inset: The scaling of the critical perturbation strength~$\lambda^*$ with the system size. The solid line is the exponential fit with $e^{-0.44 L}$.}
	\label{fig_AGP_XXX_quasi_int}
\end{figure}

We now proceed with calculating the AGP norm for the spin-$\frac{1}{2}$ XXX model weakly perturbed by an isotropic next-to-nearest neighbor interaction.
The Hamiltonian is $H(\lambda) = H_0 + V(\lambda)$, where $H_0$ is the integrable XXX Hamiltonian~(\ref{H_XXX}) and the perturbation $V(\lambda)$ is given by Eq.~(\ref{NNN_exchange}).
In this section we consider the system on a finite lattice of $L$ sites and impose periodic boundary conditions in order to retain the translation invariance of the model. For the AGP norm in Eq.~(\ref{AGP_norm}) we include only the eigenstates of $H(\lambda)$ belonging to the Hilbert space sector with zero magnetization~\cite{Sz_0_sector_note}. Accordingly, for the normalization factor in Eq.~(\ref{AGP_norm}) we take the size~${\cal D}_0$ of zero magnetization sector. Similarly, the cutoff is chosen as~$\mu = L/{\cal D}_0$. 
Then, using Eq.~(\ref{AGP_norm}) we calculate numerically the AGP norm as a function of system size for $8 \leq L \leq 20$.
Since in our case the perturbation $V(\lambda)$ is extensive, we rescale the AGP norm as $\| {\cal A}_{\lambda}\|^2 /L$.
The results are presented in Fig.~\ref{fig_AGP_XXX_quasi_int} for different values of perturbation strength~$\lambda$.
One can clearly see that the rescaled AGP norm enters the regime of exponential scaling at a certain system size-dependent critical perturbation strength~$\lambda^*(L)$. For $\lambda \lesssim \lambda^*(L)$, the (rescaled) AGP norm scaling is bounded by a polynomial in $L$, as one would expect for an integrability-preserving perturbation~\cite{pandey2020adiabatic}. Interestingly, we find that in our case the scaling is logarithmic, $\| {\cal A}_{\lambda} \|^2/L \sim \log L$. 
In the opposite case, for $ \lambda \gtrsim \lambda^*(L)$, the rescaled AGP norm scales exponentially with $L$. In this regime one has $\| {\cal A}_{\lambda} \|^2/L \sim  e^{ \kappa L}$ and our results give $\kappa = 1.62 \pm 0.05$.
As we demonstrate on the inset in Fig.~\ref{fig_AGP_XXX_quasi_int}, the critical coupling decreases exponentially with the system size and we find $\lambda^* \sim e^{- 0.44 L}$.

The fact that the AGP norm in Fig.~\ref{fig_AGP_XXX_quasi_int} scales in a drastically different way from both the integrability-preserving perturbations and the genuinely chaotic ones, supports the conjecture on the quasi-integrability of the perturbed XXX chain~\cite{kurlov2021}. Moreover, our results suggest that the AGP norm can be used as a very useful tool not only for detecting chaos but also for distinguishing the chaotic perturbations from the weak integrability-breaking ones, in the spirit of the KAM theorem. Indeed, consider again the Hamiltonian $H(\lambda) = H_0 + V(\lambda)$, where $H_0$ is integrable and $V(\lambda)$ is the perturbation. Assume that $V(\lambda)$ is known to break the integrability, which is usually easy to check. Then, in order to tell whether the integrability is strongly or weakly broken, all one needs is to calculate the AGP norm in the {\it integrability-breaking} direction~$V(\lambda)$ for sufficiently large system size and zero value of $\lambda$ [i.e. one calculates the AGP using the eigenstates of an unperturbed Hamiltonian, cf. Eq.~(\ref{AGP_norm}). If the AGP norm scales exponentially, then the perturbation~$V(\lambda)$ is chaotic and it completely breaks the integrability of $H_0$. On the contrary, if the scaling of the AGP norm at $\lambda = 0$ is bounded polynomially, then $V(\lambda)$ only weakly breaks the integrability. 

\begin{figure}[t!]
	\includegraphics[width=\columnwidth]{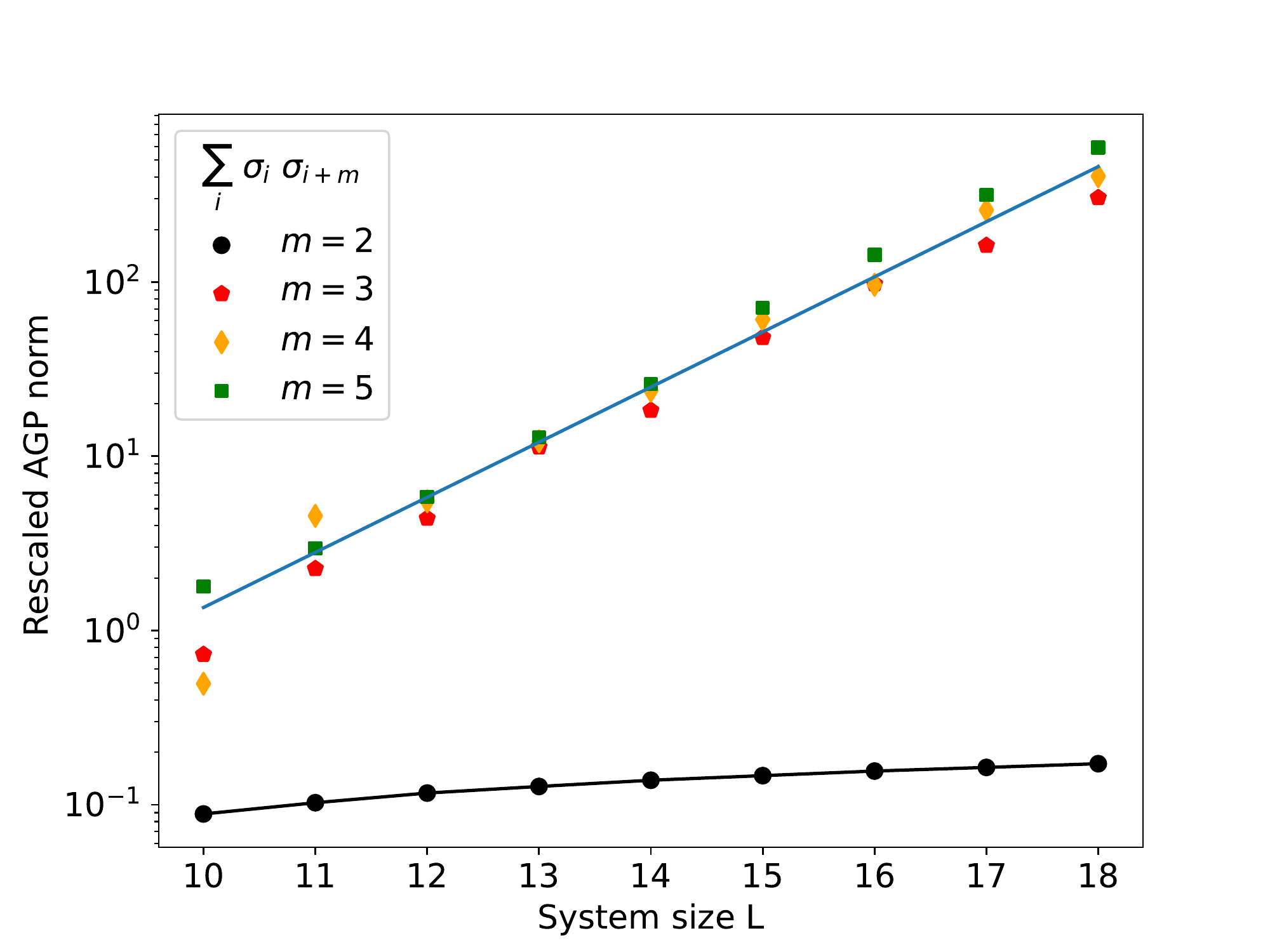}
    \caption{The rescaled AGP norm $\|{\cal A}_{\lambda}\|^2/L$ versus the system size $L$ for the XXX model perturbed by the interaction $V(\lambda) = \lambda \sum_{j} {\bs \sigma}_j \cdot {\bs \sigma}_{j+m}$ for $2 \leq m \leq 5$. In all cases the AGP norm is calculated in the direction of $V(\lambda)$ at $\lambda = 0$. For $m=2$, which corresponds to the next-to-nearest neighbor exchange in Eq.~(\ref{NNN_exchange}), the perturbation is weakly integrability-breaking, so that the rescaled AGP norm exhibits a polynomially bounded scaling (black dots). Data points for $m=2$  are the same as the $\lambda = 0$ points in Fig~\ref{fig_AGP_XXX_quasi_int}.
    For less local perturbations with $3 \leq m \leq 5$, the AGP norm scales exponentially, which indicates that these perturbations are truly chaotic as they break the integrability strongly.
    The blue solid line is the exponential fit $\|{\cal A}_{\lambda}\|^2/L \propto e^{\kappa L}$, with $\kappa \approx\text{ln}\, 2$.
    }
	\label{fig_poly_vs_exp}
\end{figure}

We illustrate this idea in Fig.~\ref{fig_poly_vs_exp}, which shows the scaling of the AGP norm for the XXX chain perturbed by the operators of the form $V(\lambda) = \lambda \sum_{j} {\bs \sigma}_j \cdot  {\bs \sigma}_{j+m}$ with $2 \leq m \leq 5$. For each $m$, we calculate the AGP at $\lambda = 0$. The results unambiguously demonstrate that the case $m=2$, which corresponds to the next-to-nearest exchange interaction, is special, since the scaling of the AGP norm is bounded polynomially. On the contrary, for less local perturbations with $3 \leq m \leq 5$ the AGP norm at $\lambda = 0$ scales exponentially with the system size. Therefore, the next-to-nearest exchange interaction breaks the integrability of the XXX model only weakly, whereas the perturbations with larger support are genuinely chaotic as they break the integrability stronger. 

 Let us finish this section with a remark. Naively, one may conclude the scaling of the AGP norm in Fig.~\ref{fig_AGP_XXX_quasi_int} strongly resembles the one presented in Fig.~3 in~Ref.~\cite{pandey2020adiabatic}. 
While it is true that the effects of integrability-breaking perturbations on the behavior of the AGP norm have already been studied in Ref.~\cite{pandey2020adiabatic},
their protocol is very different from ours. 
They consider a weakly-nonintegrable system and calculate the AGP for a perturbation that, unlike in our case, is {\it different from the one breaking the integrability.}  
Then, they find that the scaling of the AGP norm in the {\it integrable} direction also demonstrates the crossover between the regimes of polynomially bounded and exponential scalings, similar to the one in Fig.~\ref{fig_AGP_XXX_quasi_int}. 
We would like to emphasize that our protocol (where the AGP is calculated in the direction of the same perturbation that breaks the integrability) allows for a transparent physical interpretation, as demonstrated in this section.

\section{Long-range deformations and quasiconserved charges} \label{sec_quasi-int}
In this section we proceed with proving the conjecture on the quasi-integrability of the Heisenberg chain~(\ref{H_XXX}) weakly perturbed by the next-to-nearest neighbor exchange interaction~(\ref{NNN_exchange}). First, we briefly review the idea of integrability-preserving long-range deformations, discussed in Ref.~\cite{Bargheer2009}, and relate it with the AGP. Then, we show how truncating the formal series for the charges of a long-range deformed integrable spin chain leads us to a quasi-integrabile and with quasiconserved charges. We then present an explicit construction of the quasiconserved charges for the perturbed XXX chain.

\subsection{AGP and integrability-preserving deformations}

We consider the Hamiltonian $H(\lambda) = H_0 + V(\lambda)$ from Eq.~(\ref{H_tot}). The term $H_0$ is the integrable part, and one has an infinite set of mutually commuting conserved charges~$Q_{n}^{(0)}$. Then, let $\{ \ket{n_0} \}$ and $\{ \ket{n(\lambda)} \}$ be the eigenbasis of $H_0$ and $H(\lambda)$, respectively, so that the transformation $U(\lambda)$ from Eq.~(\ref{basis_rotation_U}) connects the unperturbed basis with the perturbed one. The transformation~$U(\lambda)$ is generated by the AGP~${\cal A}_{\lambda}$, as follows from Eq.~(\ref{AGP_unitary}). The AGP satisfies the operator relation~(\ref{AGP_op_rel}).

Then, following Ref.~\cite{Bargheer2009}, let us assume that the perturbation~$V(\lambda)$ does not break the integrability of the total Hamiltonian~$H(\lambda)$. In this case one has an infinite set of deformed conserved charges $Q_n(\lambda)$ that satisfy
\begin{equation} \label{deformed_charges_commutativity}
    \bigl[ H(\lambda), Q_n(\lambda) \bigr] = \bigl[ Q_m(\lambda), Q_n(\lambda) \bigr] = 0,
\end{equation}
along with $Q_n(0) = Q_{n}^{(0)}$. Keeping in mind that $H(\lambda)$ satisfies the relation~(\ref{AGP_op_rel}), and the Hamiltonian is a conserved charge itself [$H(\lambda) = Q_2(\lambda)$ by convention], let us try  a deformation similar to (\ref{AGP_op_rel}) for the other charges:
\begin{equation} \label{charges_deformatoion_def}
    i \partial_{\lambda} Q_n(\lambda) = \bigl[{\cal A}_{\lambda}, Q_n(\lambda) \bigr] - i {\cal C}_n(\lambda),
\end{equation}
where ${\cal C}_n(\lambda)$ is an operator commuting with {\it all} charges~$Q_{m}(\lambda)$. 
Denoting by ${\cal E}_{n,m}(\lambda)$ the eigenvalue of $Q_n(\lambda)$ that corresponds to the eigenstate $\ket{m(\lambda)}$, we have the spectral decomposition ${\cal C}_{n}(\lambda) = - \sum_{m} \partial_{\lambda} {\cal E}_{n,m}(\lambda) \ket{m(\lambda)}\, \bra{m(\lambda)}$, which is similar to that in Eq.~(\ref{F_op}).
On the other hand, ${\cal C}_n(\lambda)$ can be written as a linear combination of conserved charges:
\begin{equation} \label{C_n_decomposition}
    {\cal C}_{n}(\lambda) = \sum_{m} \alpha_{n,m}(\lambda) Q_{m}(\lambda).
\end{equation}
Moreover, due to the fact that the charges are defined up to an arbitrary linear transformation, the coefficients $\alpha_{n,m}(\lambda)$ in Eq.~(\ref{C_n_decomposition}) can be arbitrary functions of $\lambda$.
Note that for $n=2$ Eq.~(\ref{charges_deformatoion_def}) reduces to Eq.~(\ref{AGP_op_rel}).

Using the Jacobi identity, we immediately obtain
\begin{equation} \label{QM_QN_comm_DE}
    i \partial_{\lambda} \bigl[ Q_m(\lambda), Q_n(\lambda) \bigr] = \Bigl[ {\cal A}_{\lambda}, \bigl[ Q_m(\lambda), Q_n(\lambda) \bigr] \Bigr].
\end{equation}
Because the initial condition is $\bigl[ Q_m(0), Q_n(0)\bigr] = 0$, the solution to Eq.~(\ref{QM_QN_comm_DE}) is identically zero, i.e.~$\bigl[ Q_m(\lambda), Q_n(\lambda) \bigr] = 0$, in agreement with Eq.~(\ref{deformed_charges_commutativity}). In other words, for a set of mutually commuting unperturbed charges $Q_n^{(0)}$ there always exists a commutativity-preserving deformation~(\ref{charges_deformatoion_def}). 
Of course, this does not imply that an arbitrary perturbation~$V(\lambda)$ is integrability-preserving. Indeed, in general the deformed charges~$Q(\lambda)$ from Eq.~(\ref{charges_deformatoion_def}) at non-zero~$\lambda$ lose the {\it locality} property. 
In order for the deformed charges~$Q_n(\lambda)$ to be local,
\begin{equation} \label{Q_n_local}
    Q_n(\lambda) = \sum_{j} q_{n,j}(\lambda),
\end{equation}
the AGP~${\cal A}_{\lambda}$ in Eq.~(\ref{charges_deformatoion_def}) must belong to certain special classes of operators, as was shown in~Ref.~\cite{Bargheer2009}. Namely, ${\cal A}_{\lambda}$ can be ({\it i}) local and $\lambda$-independent; ({\it ii}) a Boost operator constructed out of one of the charges $Q_n(\lambda)$; or ({\it iii}) the so-called bilocal operator. If the AGP in Eq.~(\ref{charges_deformatoion_def}) belongs to one of these three cases, then the deformed charges~$Q_n(\lambda)$ are of the form~(\ref{Q_n_local}). 
For the sake of completeness, below we briefly discuss these three cases, before we proceed to explicitly construct the infinite family of quasi-conserved charges for the perturbed XXX chain. 

The most obvious choice for the integrability-preserving deformation is the local and $\lambda$-independent AGP, i.e.
\begin{equation} \label{AGP_local}
    {\cal A}_{\lambda}^{\text{(loc)}} = \sum_{j} A_j,
\end{equation}
where $A_{j} \neq A_j(\lambda)$ has a finite support. This is a trivial case since it corresponds to merely a  basis transformation for the unperturbed Hamiltonian~$H_0$ with the unitary operator~$U(\lambda) = e^{- i \lambda \sum_j A_j}$.
Another option for the integrability-preserving long-range deformation is the AGP of the form 
\begin{equation} \label{AGP_boost}
    {\cal A}_{\lambda}^{\text{(boost)}} \propto {\cal B}[Q_{n}(\lambda)] = \frac{1}{2i}\sum_{j} j q_{n,j}(\lambda),
\end{equation}
i.e. the AGP is the boost operator of the $n$th conserved charge~(\ref{Q_n_local}).
Finally, one can show~\cite{Bargheer2009} that taking the so-called {\it bilocal} operator for the AGP also yields an integrability-preserving deformation:
\begin{multline} \label{AGP_bilocal}
    {\cal A}_{\lambda}^{\text{(biloc)}} \propto \left[ Q_{m}(\lambda) | Q_{n}(\lambda) \right] \\ 
    \equiv \frac{1}{2} \sum_{j} \{ q_{m,j}(\lambda), q_{n,j}(\lambda) \} + \sum_{i < j} \{ q_{m,i}, q_{n,j} \},
\end{multline}
where $\{ \cdot, \cdot \}$ is the anticommutator. It is straightforward to check the commutator in Eq.~(\ref{charges_deformatoion_def}) results in a local operator if the AGP is given by Eq~(\ref{AGP_local}), (\ref{AGP_boost}), or (\ref{AGP_bilocal}).

For a given AGP, one can easily solve Eq.~(\ref{charges_deformatoion_def}) perturbatively in $\lambda$. Indeed, the long-range Hamiltonian~$H(\lambda) = H_0 + V(\lambda)$ can be written as a formal power series in $\lambda$ as
\begin{equation} \label{H_lambda_series}
    H(\lambda) = H_0 + \sum_{k=1}^{+\infty} \lambda^k V_k.    
\end{equation}
The deformation~$V(\lambda)$ is generated by some AGP~${\cal A}_{\lambda}$.
Then, for this AGP and the corresponding deformed charges~$Q_n(\lambda)$ we write similar formal expansions:
\begin{gather} 
    {\cal A}_{\lambda} = \sum_{k = 0}^{+ \infty} \lambda^k {\cal A}^{(k)}, \label{A_lambda_series} \\
    Q_n(\lambda) = \sum_{k=0}^{+\infty} \lambda^k Q_n^{(k)}. \label{Q_lambda_series} 
\end{gather}
Then, from Eq.~(\ref{charges_deformatoion_def}) we immediately obtain
\begin{multline} \label{quasi_charges_iteratively}
    \sum_{k=0}^{+ \infty} i (k+1) \lambda^{k} Q_n^{(k+1)} = \sum_{p,r=0}^{+ \infty} \lambda^{p+r} \left[ {\cal A}^{(p)}, Q_n^{(r)} \right] \\ 
    - i \sum_{p,r=0}^{+ \infty} \frac{\lambda^{p+r}}{p!} \sum_{m} \alpha_{n,m}^{(p)}(0) Q_{m}^{(r)},
\end{multline}
which allows one to construct $Q_n(\lambda)$ to the desired order in $\lambda$ iteratively. 
Let us emphasize that $Q_n(\lambda)$ commute with each other and $H(\lambda)$ only if one takes into account the complete infinite series in Eqs.~(\ref{H_lambda_series}), (\ref{A_lambda_series}), and (\ref{Q_lambda_series}). Truncating the series would lead to {\it approximate} conservation laws, which we are going to use below.

We also note that the construction of long-range deformed integrable Hamiltonians outlined above is deeply connected with the so-called $T \bar T$-deformations, well known in the field theory literature, see e.g. Ref.~\cite{Pozsgay2020}.

\subsection{Quasiconserved charges in the perturbed XXX model}

We now return to the Heisenberg model $H_0$ in Eq.~(\ref{H_XXX}) perturbed by the next-to-nearest exchange interaction~$V(\lambda) = \lambda V_1$, where 
\begin{equation} \label{V1_NNN_exchange}
    V_1 = \sum_{j} {\bs \sigma}_j \cdot {\bs \sigma}_{j+2},
\end{equation}
as described in detail in Sec.~\ref{sec_model}. Our aim is to prove that the perturbed XXX Hamiltonian $H(\lambda) = H_0 + \lambda V_1$ has as many quasi-conserved charges as there are exactly conserved ones for the unperturbed XXX model, a conjecture put forward in Ref.~\cite{kurlov2021}.

The idea of the proof is extremely simple. Let us consider an integrability-preserving long-range deformation of ~$H_0$, generated by some appropriate AGP~${\cal A}_{\lambda}$. Further, assume that one can choose~${\cal A}_{\lambda}$ in such a way that to the first order in $\lambda$ the solution to Eq.~(\ref{AGP_op_rel}) is nothing else than~$H(\lambda) = H_0 + \lambda V_1$.
Keeping in mind that $H(\lambda) = Q_2(\lambda)$, so that $H_0 = Q_2^{(0)}$ and $V_1 = Q_2^{(1)}$, one can equivalently work with Eq.~(\ref{quasi_charges_iteratively}), after setting there $n=2$ and truncating both sides to the {\it zeroth} order in $\lambda$ [one power is lost after taking the derivative in Eq.~(\ref{AGP_op_rel})]. This yields
\begin{equation} \label{AGP_quasi_integrable}
    V_1 = - i\bigl[ {\cal A}^{(0)}, H_0 \bigr] - \sum_{m} \alpha_m Q_{m}^{(0)},
\end{equation}
where we denoted $\alpha_m \equiv \alpha_{2,m}(0)$. Thus, our goal is to find the operator~${\cal A}^{(0)}$ and the coefficients $\alpha_{m}$ such that Eq.~(\ref{AGP_quasi_integrable}) is satisfied for a given perturbation~$V_1$.
If this is possible, one can immediately construct {\it all} quasi-conserved charges by truncating the series in Eq.~(\ref{Q_lambda_series}) as 
\begin{equation}
    \tilde Q_n(\lambda) = Q_n^{(0)} + \lambda Q_n^{(1)},
\end{equation}
and then solving for ~$Q_n^{(1)}$ from~Eq.~(\ref{quasi_charges_iteratively}), truncated to the zeroth order in $\lambda$:
\begin{equation} \label{Qn1_quasi_integrable}
    Q_n^{(1)} = - i \bigl[ {\cal A}^{(0)}, Q_n^{(0)} \bigr] - \sum_{m} \alpha_{n, m} Q_{m}^{(0)},
\end{equation}
with $\alpha_{n, m} \equiv \alpha_{n, m}(0)$ for brevity.
The quasi-conserved charges~$\tilde Q_n$ commute with each other and the Hamiltonian $H(\lambda) = H_0 + \lambda V_1$ with the accuracy $O(\lambda^2)$ {\it by construction}.

We are now finally in the position to demonstrate that the algorithm outlined above can be performed for the perturbed XXX chain. 
Indeed, consider the zeroth order AGP of the form
\begin{equation} \label{AGP_boost_Q3}
    {\cal A}^{(0)} = i {\cal B}\bigl[ Q_3^{(0)} \bigr] = \frac{1}{2}\sum_{j} j \left( {\bs \sigma}_j \times {\bs \sigma}_{j+1} \right) \cdot {\bs \sigma}_{j+2},
\end{equation}
which is nothing other than the boost operator constructed from the conserved charge $Q_3^{(0)}$ of the Heisenberg Hamiltonian, see Eq.~(\ref{Q_3_XXX}). It is straightforward to check that for the Heisenberg Hamiltonian $H_0$ from Eq.~(\ref{H_XXX}) one has
\begin{equation}
    - i \bigl[ {\cal A}^{(0)}, H_0 \bigr] = V_1 - 2 H_0 - Q_4^{(0)},
\end{equation}
where $Q_4^{(0)}$ is the conserved charge of the Heisenberg model, explicitly given by Eq.~(\ref{Q_4_XXX}).
Therefore, given the perturbation~$V_1$ from Eq.~ (\ref{V1_NNN_exchange}) and the AGP from Eq.~(\ref{AGP_boost_Q3}), one can satisfy Eq~(\ref{AGP_quasi_integrable}) if we set the coefficients $\alpha_m = 2 \delta_{m,2} + \delta_{m,4}$.
We then immediately obtain the infinite tower of quasi-conserved charges~$\tilde Q_n(\lambda)$ for the perturbed XXX model, given in terms of the charges~$Q_n^{(0)}$ of the unperturbed Hamiltonian. Explicitly, from Eq.~(\ref{Qn1_quasi_integrable}) for any $n \geq 3$ we have
\begin{equation}
    \tilde Q_n(\lambda) = Q_n^{(0)} - \lambda \sum_{m} \alpha_{n,m} Q_m^{(0)} + \lambda \left[ {\cal B}\bigl[ Q_3^{(0)} \bigr], Q_n^{(0)} \right],
\end{equation}
where $\alpha_{n,m}$ are arbitrary real numbers. Therefore, we have proven the conjecture of Ref.~\cite{kurlov2021} on the quasi-integrability of the perturbed XXX chain. In the same way one can easily construct other quasi-integrable models and their quasi conserved charges. It is also straightforward to generalize this procedure to an arbitrary order in~$\lambda$. 

For completeness, in App.~\ref{app_XX_quasicharges} we demonstrate an example of a weakly-nonintegrable model (isotropic XY chain perturbed by the next-to-nearest XY interaction), whose quasiconserved charges are constructed with the help of the AGP from the class of bilocal operators.

\section{Conclusions} \label{sec_conclusions}

In this work, we first have studied weak integrability breaking in the spin-$\frac{1}{2}$ Heisenberg chain (XXX model) perturbed by the isotropic next-to-nearest neighbor exchange interaction. We have shown that in this model the AGP norm scales with the system size in a distinct way, different from both the polynomially bounded scaling characteristic of integrability-preserving perturbations and the exponential scaling for the chaotic ones. Instead, for the perturbed XXX model we find that the (rescaled) AGP norm exhibits a sharp crossover from the regime of polynomially bounded to the exponential scaling. The crossover occurs at a critical perturbation strength that is exponentially small in the system size, and we have~$\lambda^* \sim e^{- 0.44 L}$. 
In the regime of exponential scaling, i.e. for $\lambda \gtrsim \lambda^*$, for the rescaled AGP norm we have found $\| {\cal A}_{\lambda} \|^2 / L \propto e^{\kappa L}$, where $\kappa = 1.62\pm0.05$. On the contrary, for $\lambda \lesssim \lambda^*$ the scaling of the AGP norm is bounded polynomially, just like it is for intgerability-preserving perturbations. These findings strongly support the conjectured quasi-integrability of the perturbed XXX chain made in Ref.~\cite{kurlov2021}.

In the second part of this work we have presented an analytic proof of this conjecture. Using the algebraic methods developed in Ref.~\cite{Bargheer2009}, we explicitly constructed an infinite tower of quasi-conserved charges, which commute with one another and the Hamiltonian of the perturbed XXX model with the accuracy~$O(\lambda^2)$. The main idea of our proof is based on the fact that the perturbed XXX model can be viewed as the truncation to the first order in $\lambda$ of an integrable long-range spin-chain with the Hamiltonian~$H(\lambda) = H_0 + \sum_{k=1}^{+\infty}\lambda^k V_k$, where the operators $V_k$ have an increasingly large support.
The generator of this long range is nothing other than the AGP, satisfying certain additional requirements needed to preserve locality. In order to generate the deformation that to the first order in $\lambda$ gives the next-to-nearest neighbor exchange interaction, one has to take the AGP whose value at $\lambda = 0$ yields the boost operator constructed from the third conserved charge of the unperturbed XXX model.

Together, our results demonstrate that the AGP norm can be used not only to detect the onset of chaos, but also a useful tool in distinguishing different types of integrability-breaking perturbations, i.e. the truly chaotic perturbations from those that only weakly break the integrability. In the latter case the system possesses a macroscopic number of quasi-conserved charges, that can be found using the approach discussed in Sec.~\ref{sec_quasi-int}. In order to tell whether a given perturbation $V(\lambda)$ is chaotic or weakly integrability-breaking, one simply needs to calculate the AGP in the direction of $V(\lambda)$ at $\lambda = 0$. Then, the (rescaled) AGP norm scales exponentially for the chaotic perturbations, whereas for weakly integrability-breaking ones the scaling is bounded polynomially.
We expect that our findings can be useful for the studies of transport properties in weakly-nonintegrable models, see e.g.~\cite{jung2006transport, DeNardis2021, Bastianello2021}.
It would be interesting to further investigate the spectral properties of the weakly perturbed XXX chain in the regime of $\lambda \lesssim \lambda^*(L)$, and other quasi-integrable models in similar settings. For instance, from the results of Refs.~\cite{szasz2021weak, McLoughlin2022} one can expect that in this regime the level spacing statistics should not deviate significantly from the Poissonian, whereas for $\lambda \gtrsim \lambda^*(L)$ it should gradually deform into the Wigner-Dyson statistics.
We leave these questions for future work. 

Before we finish, let us make a general remark~\cite{note_Referee} on the spectral-based criteria to quantum chaos. It is important to keep in mind that the many-body eigenstates are not observables, and it takes an exponentially large (in the system size) time to resolve the individual eigenstates corresponding to exponentially close energies. As a result, the properties of the eigenstates can also be exponential sensitive to perturbations, which makes detecting chaos much harder. For instance, it is well known that some {\it non-chaotic} systems can have Wigner-Dyson-like level spacing distributions, such as quadratic systems with two bosonic modes~\cite{Benet2003}, or spin systems with a single impurity~\cite{Santos2004}. 
Finally, a common drawback of all spectral-based measures, including the AGP, is that they are limited to relatively small system sizes, amenable to exact diagonalization.

{\it Note added}. While finishing the manuscript, we have learned that the authors of Ref.~\cite{Surace2023} obtained similar results on the existence of quasi-conserved charges for weakly nonintegrable Hamiltonians.
As far as our studies overlap, our results are in agreement with each other.
Key differences between Ref.~\cite{Surace2023} and our work are the following: (i) we additionally investigate the scaling of the AGP norm and find that for a weakly-nonintegrable Heisenberg model it behaves in a distinct way; (ii) the authors of Ref.~\cite{Surace2023} extend the proof of quasi-integrability to Hamiltonians with higher order perturbations and illustrate the procedure with a number of different models.

\section*{Acknowledgements}
The numerical computations in this work were performed using QuSpin~\cite{QuSpin_I,QuSpin_II}.
We acknowledge useful discussions with Igor Aleiner, Boris Altshuler, Jacopo de Nardis, Anatoli Polkovnikov, and Gora Shlyapnikov. We thank Piotr Sierant and Dario Rosa for drawing our attention to Refs.~\cite{Sierant2019, Sierant2020, Sierant2022}  and Ref.~\cite{Nandy2022}, respectively.
We are grateful to an anonymous referee for very useful comments and for drawing our attention to Refs.~\cite{Benet2003, Santos2004}.
The work of VG is part of the DeltaITP consortium, a program of the Netherlands Organization for Scientific Research (NWO) funded by the Dutch Ministry of Education, Culture and Science (OCW). VG is also partially supported by RSF 19-71-10092.
The work of AT was supported by the ERC Starting Grant 101042293 (HEPIQ). RS acknowledges support from Slovenian Research Agency (ARRS) - research programme P1-0402. 

\appendix
\section{Quasi-conserved charges of the perturbed isotropic XY chain} \label{app_XX_quasicharges}

In this appendix we discuss the construction of quasiconserved charges for the isotropic XY chain weakly perturbed by the next-to-nearest XY interaction, studied in Ref.~\cite{bahovadinov2022many}.
The Hamiltonian is given by
\begin{equation}\label{XX_perturbed}
    H(\lambda) = H_0 + \lambda V_1,
\end{equation}
where $H_0$ is the integrable part
\begin{equation}
    H_0 =  \sum_{j} \left( \sigma_{j}^{x}\sigma_{j+1}^{x} + \sigma_{j}^{y}\sigma_{j+1}^{y}  \right),
\end{equation}
and $V_1$ is the perturbation of strength $\lambda \ll 1$, which reads
\begin{equation} \label{XX_V1}
    V_1 =  \sum_{j} \left( \sigma_{j}^{x}\sigma_{j+2}^{x} + \sigma_{j}^{y}\sigma_{j+2}^{y}  \right).
\end{equation}
At $\lambda = 0$ the model is integrable (it can be mapped onto free fermions) and it has two families of conserved quantities \cite{grabowski1995structure}. 
Explicitly, the first family is given by
\begin{equation}\label{family1}
    Q_n^{(0)} = \begin{cases}
    \sum\limits_{j} \left( e_{n,j}^{xx} + e_{n,j}^{yy}\right), \qquad  \text{$n$ even};
    \\
    \sum\limits_{j} \left( e_{n,j}^{xy} - e_{n,j}^{yx}\right), \qquad \text{$n$ odd},
    \end{cases}
\end{equation}
and the second family can be written as
\begin{equation}\label{family2}
    I_n^{(0)} = \begin{cases}
     \sum\limits_{j} \left( e_{n,j}^{xy} - e_{n,j}^{yx}\right), \qquad \text{$n$ even};
    \\
    \sum\limits_{j} \left( e_{n,j}^{xx} + e_{n,j}^{yy}\right), \qquad \text{$n$ odd},
    \end{cases}
\end{equation}
where $n \geq 2$. In Eqs.~(\ref{family1}) and~(\ref{family2}) we introduced the operators
\begin{equation}
    e_{n,j}^{\alpha \beta } = \sigma_{j}^{\alpha} \sigma_{j+1}^{z} ... \sigma_{j+n-2}^{z} \sigma_{j+n-1}^{\beta}.
\end{equation}
 For instance, from Eq.~(\ref{family1}) with $n=2$ one has
\begin{equation}
    Q_2^{(0)} = \sum_{j} \left( \sigma_{j}^{x}\sigma_{j+1}^{x} + \sigma_{j}^{y}\sigma_{j+1}^{y}  \right)
\end{equation}
which is simply the isotropic XY Hamiltonian $H_0$ itself, and Eq.~(\ref{family2}) for $n=2$ gives
\begin{equation}
    I_2^{(0)} = \sum_j \left( \sigma_{j}^{x}\sigma_{j+1}^{y} -\sigma_{j}^{y}\sigma_{j+1}^{x}  \right),
\end{equation}
which is the Dzyaloshinski-Moriya interaction.
The conserved charges $Q_n^{(0)}$ are invariant under the parity transformation $\sigma_{j}^{\alpha} \rightarrow - \sigma_{j}^{\alpha}$, whereas the charges $I_n^{(0)}$ change their sign.

As was shown in Ref.~\cite{bahovadinov2022many} the perturbed Hamiltonian~(\ref{XX_perturbed}) possesses a quasi-conserved quantity that commutes with~$H(\lambda)$ with the accuracy~$O(\lambda^2)$.
Let us try to construct the AGP that generates this quasiconserved charge and try to construct more of them.
In analogy with the perturbed XXX model, discussed in Sec.~\ref{sec_quasi-int}, one  could try to use the AGP proportional to the boost operator constructed from~$Q_3^{(0)}$ or $I_3^{(0)}$.
However, it turns out that for the isotropic XY model the boost operators ${\cal B}[Q_n^{(0)}]$  and ${\cal B}[I_n^{(0)}]$ with {\it any} $n$ simply generate other charges from Eqs.~(\ref{family1}) and (\ref{family2}). For this reason, one can not generate a nontrivial long-range deformation of the isotropic $XY$ chain using only boost operators.

Therefore, we have to look for the AGP in the class of bilocal operators.
Let us try the following one:
\begin{equation}
    \mathcal{A}_{\lambda} = \frac{1}{4} [ S_{z}  | I_{2}(\lambda)],
\end{equation}
where $S_{z} = \sum_{j} \sigma_{j}^{z}$ is the $z$-projection of total spin and the factor of $\frac{1}{4}$ is included for later convenience. To zeroth order in $\lambda$, the explicit expression for the AGP reads
\begin{multline} \label{perturbed_XX_AGP_0}
    {\cal A}^{(0)} =  \frac{1}{4} [ \sigma^{z}  | I_{2}^{(0)} ] \\
    =  \frac{1}{2} \sum_{j} \sum_{r > 0} \sigma_{j}^{z} ( \sigma_{j+r}^{x} \sigma_{j+r+1}^{y} - \sigma_{j+r}^{y} \sigma_{j+r+1}^{x}).
\end{multline} 
Then, using Eg.~(\ref{AGP_quasi_integrable}) we obtain 
\begin{multline} \label{XX_correction_def}
V_1  = \sum_{j} \sum_{\alpha = x,y} \sigma_{j}^{\alpha}\sigma_{j+2}^{\alpha} + 2 \sum_{j} \sigma_{j}^{z} \sigma_{j+1}^{z} \\
- \sum_{m} \left( \beta_m Q_m^{(0)} + \gamma_m I_m^{(0)} \right),
\end{multline}
where we took into account that the unperturbed model has two families of conserved charges.
The first term on the right hand side of Eq.~(\ref{XX_correction_def}) is precisely the perturbation~$V_1$ from Eq.~(\ref{XX_V1}). However, the second term is the nearest neighbor Ising interaction, and it is clearly impossible to cancel it using the charges $Q_m^{(0)}$ and $I_m^{(0)}$.

The trick here is to add a correction to the AGP, which would eliminate the Ising interaction from Eq.~(\ref{XX_correction_def}). One immediately observes that this correction is exactly the AGP that generates the deformation of the isotropic XY chain into the XXZ chain with the Hamiltonian
\begin{equation} \label{H_XXZ}
    H_{\text{XXZ}} = \sum_{j} \left( \sigma^x_j \sigma^x_{j+1} + \sigma^y_j \sigma^y_{j+1} 
    - 2 \lambda \sigma^z_j \sigma^z_{j+1} \right).
\end{equation}
Taking into account Eq.~(\ref{H_XXZ}), let us rewrite the perturbation~(\ref{XX_V1}) as
\begin{equation} \label{V1_XXZ}
    V_1 = - i \bigl[ {\cal A}^{(0)}, H_0 \bigr] + \partial_{\lambda} H_{\text{XXZ}}.
\end{equation}
Keeping in mind that the Hamiltonian $H(\lambda) = H_0 + \lambda V_1$ is a (quasiconserved) charge itself, one can see that the remaining charges are constructed in a way similar to Eq.~(\ref{V1_XXZ}). Thus, we write
\begin{equation} 
    Q_n^{(1)} = - i [ {\cal A}^{(0)} , Q_n^{(0)}] + \left( \partial_{\lambda} Q_n^{\text{XXZ}} \right) \bigr|_{\lambda = 0},
\end{equation}
where ${\cal A}^{(0)}$ is given by Eq.~(\ref{perturbed_XX_AGP_0}), $Q_n^{\text{XXZ}}$ is the $n$th conserved charge of the XXZ model~(\ref{H_XXZ}), and we only keep the terms in $Q_n^{\text{XXZ}}$ that are linear in~$\lambda$.
The conserved charges $Q_n^{\text{XXZ}}$ can be genrated with the help of the boost operator as 
\begin{equation}
    Q_{n+1}^{\text{XXZ}} \propto  \bigl[ {\cal B}[H_{\text{XXZ}}], Q_n^{\text{XXZ}} \bigr].
\end{equation}
Then, the family of charges in Eq.~(\ref{family1}) gets deformed as
\begin{equation} \label{XX_NNN_quasicharges}
    \tilde Q_n(\lambda) = Q_n^{(0)} + \lambda Q_n^{(1)}.
\end{equation}
Note that the second family of charges, given by Eq.~(\ref{family2}), is destroyed by the perturbation.
 It is straightforward to check that the quasicoserved charges~(\ref{XX_NNN_quasicharges}) commute with each other and the perturbed isotropic XY chain~(\ref{XX_perturbed}) with the accuracy~$O(\lambda^2)$.

\bibliography{references}

\begin{thebibliography}{84}
\bibitem [{not({\natexlab{a}})}]{note_quasi_integrability}%
  \BibitemOpen
  \href@noop {} {} ({\natexlab{a}}),\ \bibinfo {note} {{Let} us emphasize that
  by {\it quasi-integrability} we mean the existence of approximately conserved
  (quasi-conserved). This should not be confused with the {\it
  pseudo-integrability}, which has a different meaning and denotes a classical
  system (i.e. a polygonal billiard) that possesses conserved quantities, but
  whose invariant manifold is not a torus.}\BibitemShut {Stop}%
\bibitem [{Sz_()}]{Sz_0_sector_note}%
  \BibitemOpen
  \href@noop {} {}\bibinfo {note} {Strictly speaking, we use the sector with
  $N_{\uparrow} = \lfloor L/2 \rfloor$ up spins and $N_{\downarrow} = L -
  N_{\uparrow}$ down spins, where $\lfloor \cdot \rfloor$ denotes the floor
  function. Then, for the total magnetization $S_z = (N_{\uparrow} -
  N_{\downarrow} )/2$ we obtain $S_z = 0$ ($S_z = -1/2$) if $L$ is even (odd).
  For our purposes this difference between the case of even and odd $L$ can be
  ignored.}\BibitemShut {Stop}%
\end{thebibliography}

\end{document}